\documentclass[twocolumn]{aastex62}

\received{2020 April 20}
\accepted{2020 May 18}

\shorttitle{Grain growth in newly discovered young eruptive stars}
\shortauthors{K\'osp\'al et al.}

\begin{document}

\title{Grain growth in newly discovered young eruptive stars\footnote{Based on observations collected at the European Southern Observatory under ESO programs 097.C-0292, 098.C-0529, and 0102.C-0310.}}

\author[0000-0001-7157-6275]{\'A. K\'osp\'al}
  \affiliation{Konkoly Observatory, Research Centre for Astronomy and Earth Sciences, Konkoly-Thege Mikl\'os \'ut 15-17, 1121 Budapest, Hungary}
  \affiliation{Max Planck Institute for Astronomy, K\"onigstuhl 17, 69117 Heidelberg, Germany}
  \affiliation{ELTE E\"otv\"os Lor\'and University, Institute of Physics, P\'azm\'any P\'eter s\'et\'any 1/A, 1117 Budapest, Hungary}

\author[0000-0001-6015-646X]{P. \'Abrah\'am}
  \affiliation{Konkoly Observatory, Research Centre for Astronomy and Earth Sciences, Konkoly-Thege Mikl\'os \'ut 15-17, 1121 Budapest, Hungary}
  \affiliation{ELTE E\"otv\"os Lor\'and University, Institute of Physics, P\'azm\'any P\'eter s\'et\'any 1/A, 1117 Budapest, Hungary}

\author{A. Carmona}
  \affiliation{Institute de Planetologie et Astrophysique de Grenoble, Universit\'e Grenoble Alpes, CNRS, IPAG, F-38000 Grenoble, France}

\author[0000-0003-2835-1729]{L. Chen}
  \affiliation{Konkoly Observatory, Research Centre for Astronomy and Earth Sciences, Konkoly-Thege Mikl\'os \'ut 15-17, 1121 Budapest, Hungary}

\author[0000-0003-1665-5709]{J. D. Green}
  \affiliation{Space Telescope Science Institute, 3700 San Martin Dr., Baltimore, MD 21218, USA}

\author[0000-0002-2190-3108]{R. van Boekel}
  \affiliation{Max Planck Institute for Astronomy, K\"onigstuhl 17, 69117 Heidelberg, Germany}

\author[0000-0001-8445-0444]{J. A. White}
  \affiliation{Konkoly Observatory, Research Centre for Astronomy and Earth Sciences, Konkoly-Thege Mikl\'os \'ut 15-17, 1121 Budapest, Hungary}

  
\begin{abstract}
FU Orionis-type stars are young stellar objects showing large  outbursts due to highly enhanced accretion from the circumstellar disk onto the protostar. FUor-type outbursts happen in a wide variety of sources from the very embedded ones to those with almost no sign of extended emission beyond the disk. The subsequent eruptions might gradually clear up the obscuring envelope material and drive the protostar on its way to become a disk-only T~Tauri star. We used VLT/VISIR to obtain the first spectra that cover the $8-13\,\mu$m mid-infrared wavelength range in low-resolution of five recently discovered FUors. Four objects from our sample show the 10$\,\mu$m silicate feature in emission. We study the shape and strength of the silicate feature in these objects and find that they mostly contain large amorphous grains, suggesting that large grains are typically not settled to the midplane in FUor disks. This is a general characteristic of FUors, as opposed to regular T Tauri-type stars whose disks display anything from pristine small grains to significant grain growth. We classify our targets by determining whether the silicate feature is in emission or in absorption, and confront them with the evolutionary scenarios on the dispersal of the envelopes around young stars. In our sample, all Class~II objects exhibit silicate emission, while for Class~I objects, the appearance of the feature in emission or absorption depends on the viewing angle with respect to the outflow cavity. This highlights the importance of geometric effects when interpreting the silicate feature.
\end{abstract}



\section{Introduction}
\label{sec:intro}

FU\,Orionis-type objects (FUors) form a small group of pre-main sequence stars characterized by large outbursts in visible and infrared light. The brightenings are attributed to increased accretion from the circumstellar disk onto the protostar \citep{audard2014}. Enhanced accretion is often accompanied by enhanced mass loss: many FUors drive collimated jets or broad molecular outflows \citep{hk96}. The importance of FUors stems from the fact that episodic accretion is often invoked as an explanation for the luminosity problem, i.e., that protostars are typically one or two orders of magnitude fainter than theoretically expected \citep{dunham2013}. If so, then all young, Sun-like stars should go through repeated accretion outbursts.

Regarding their circumstellar structure, FUors form a heterogeneous group, but the observed differences are far from being random. In the beginning of their evolution, protostars are deeply embedded in envelopes. During the repeated outbursts, however, more and more of the envelope material is accreted and/or blown away \citep{takami2018}. Therefore, it is likely that the period when FUor outbursts happen corresponds to the important transition phase when the embedded protostar gradually clears away its enshrouding envelope to become a Class~II T~Tauri star. The observed differences among individual FUors point to an evolutionary sequence within this process.

Based on the appearance of the 10$\,\mu$m silicate feature, \citet{quanz2007c} proposed a classification scheme for FUors, suggesting that objects showing the feature in absorption are younger, still embedded in a circumstellar envelope; while objects showing the silicate band in emission are more evolved, with direct view on the surface layer of the accretion disk. A similar evolutionary sequence was outlined by \citet{green2006} based on the amount of far-infrared excess, and by \citet{kospal2017a} based on the millimeter CO line flux.

In the past ten years, several new young outbursting stars were discovered. While their light curves and optical/near-infrared spectral properties make them {\it bona fide} FUors, their status in the evolutionary sequence outlined above is not clear, due to the lack of mid-infrared spectroscopy, and often due to the lack of far-infrared photometry as well. In this paper, we place these new FUors into the evolutionary scheme of \citet{quanz2007c} and constrain the dust grain properties by observing their 10$\,\mu$m silicate feature for the first time.


\section{Target selection}
\label{sec:target}

By 2009, a large fraction of FUors known at that time had been observed with the Infrared Spectrograph (IRS) of the Spitzer Space Telescope. Until the James Webb Space Telescope (JWST) is launched, we can obtain mid-infrared spectra only with ground-based instrumentation. For the present study, we selected five FUors discovered in the last decade, that are observable from Paranal Observatory, and for which no 10$\,\mu$m spectra had ever been published before.

V582~Aur was first observed photometrically and spectroscopically by \citet{semkov2011}. Its brightening started between 1984 and 1985, it reached its maximum in 1986, and it is still close to maximum brightness \citep{semkov2013}. Its flat spectral energy distribution (SED) and millimeter observations indicate a significant amount of circumstellar material, including a massive envelope \citep{abraham2018,zsidi2019}.

V723~Car brightened by 3\,mag in the $K$ band between 2000 and 2002 \citep{tapia2015}. Its SED indicates that it is a Class~I source.

V899~Mon was discovered by the Catalina Real-time Transient Survey \citep{wils2009}. According to \citet{ninan2015}, it brightened by 3.5\,mag in the $R$ band between 1989 and 2010. It is a flat-spectrum or an early Class~II young stellar object.

V900~Mon was discovered by an amateur astronomer \citep{thommes2011}, and characterized in detail by \citet{reipurth2012} and \citet{takami2019}. It brightened by 4\,mag at optical wavelengths some time between 1953 and 2009, and its SED suggests that it is a Class~I protostar with a massive cool envelope.

V960~Mon used to be a normal young, low-mass pre-main sequence star until 2014, when it suddenly brightened by 3\,mag at optical wavelengths \citep{maehara2014, kospal2015, hackstein2015}. Its spectral energy distribution suggests that it is a Class~II object, with a remnant envelope.


\section{Observations and data reduction}
\label{sec:obs}

We collected data for our targets over several semesters with ESO's Very Large Telescope (VLT) at Paranal Observatory in Chile. Tab.~\ref{tab:log} summarizes the details of our observations. We obtained low resolution ($R \sim 350$) 8--13$\,\mu$m spectra with the VLT spectrometer and imager for the mid-infrared \citep[VISIR,][]{visir}, installed on the Melipal (UT3) telescope. We used a  1$''$ wide slit and employed an ABBA nodding pattern along the slit. We concatenated the science measurements with calibrator observations to enable precise telluric correction and flux calibration. In addition, we  used calibrator data from other programs that were observed during the same night as our science targets. The calibrator stars have spectral types of A1 (Sirius), G9.5 ($\alpha$~Mon) and K1--5 (HD~26311, HD~26967, HD~37984, HD~59381, HD~75691, and HD~169916). They were all detected at very high signal-to-noise ratio (between 150 and 9000).

For basic data reduction and extraction of the spectra, we ran the ESO VISIR spectroscopic pipeline in the EsoRex environment version 3.12.3. In order to correct for telluric features, we divided the target spectrum by a spectrum derived via interpolation between the two standard star measurements, one obtained at higher and one at lower airmass than the science target. Flux calibration was carried out by multiplying by the model spectra of the standard stars. The uncertainty on the absolute flux calibration can be as large as 20\%.

\begin{table*}[]
\centering
\caption{Log of our VLT/VISIR observations.}\label{tab:log}
\begin{tabular}{lllllll}
\hline
\hline
Object   & Distance\tablenote{References: V582 Aur: \citet{kun2017}; V723 Car: \citet{tapia2015}; V900 Mon: \citet{reipurth2012}; V899 Mon and V960 Mon: \citet{bailerjones2018}.} & Airmass      & Calibrators         & ESO run & Date & Exp. time\\
\hline
V582 Aur & 1320 pc & 1.48 -- 1.45 & HD 26311, HD 26967  & 0102.C-0310(A) & 2018 Oct 13/14    & 1500\,s \\
V723 Car & 2500 pc & 1.27 -- 1.32 & HD 48915, HD 169916 & 097.C-0292(D)  & 2016 Apr 30/May 1 & 1600\,s \\
V899 Mon & 769 pc  & 1.06 -- 1.09 & HD 48915, HD 37984  & 0102.C-0310(D) & 2018 Nov 12/13    & 1400\,s \\
V900 Mon & 1100 pc & 2.10 -- 2.57 & HD 48915, HD 75691  & 097.C-0292(C)  & 2016 May 15/16    & 1300\,s \\
         &         & 1.14 -- 1.21 & HD 48915, HD 61935  & 098.C-0529(C)  & 2016 Dec 20/21    & 1100\,s \\
V960 Mon & 1574 pc & 1.67 -- 2.30 & HD 48915, HD 169916 & 097.C-0292(B)  & 2016 Apr 30/May 1 & 2300\,s \\
         &         & 1.07 -- 1.12 & HD 48915, HD 59381  & 0102.C-0310(B) & 2018 Dec 2/3      & 2200\,s \\
\hline
\end{tabular}
\end{table*}


\section{Results and Analysis}
\label{sec:res}

\begin{figure}
   \includegraphics[width=\columnwidth]{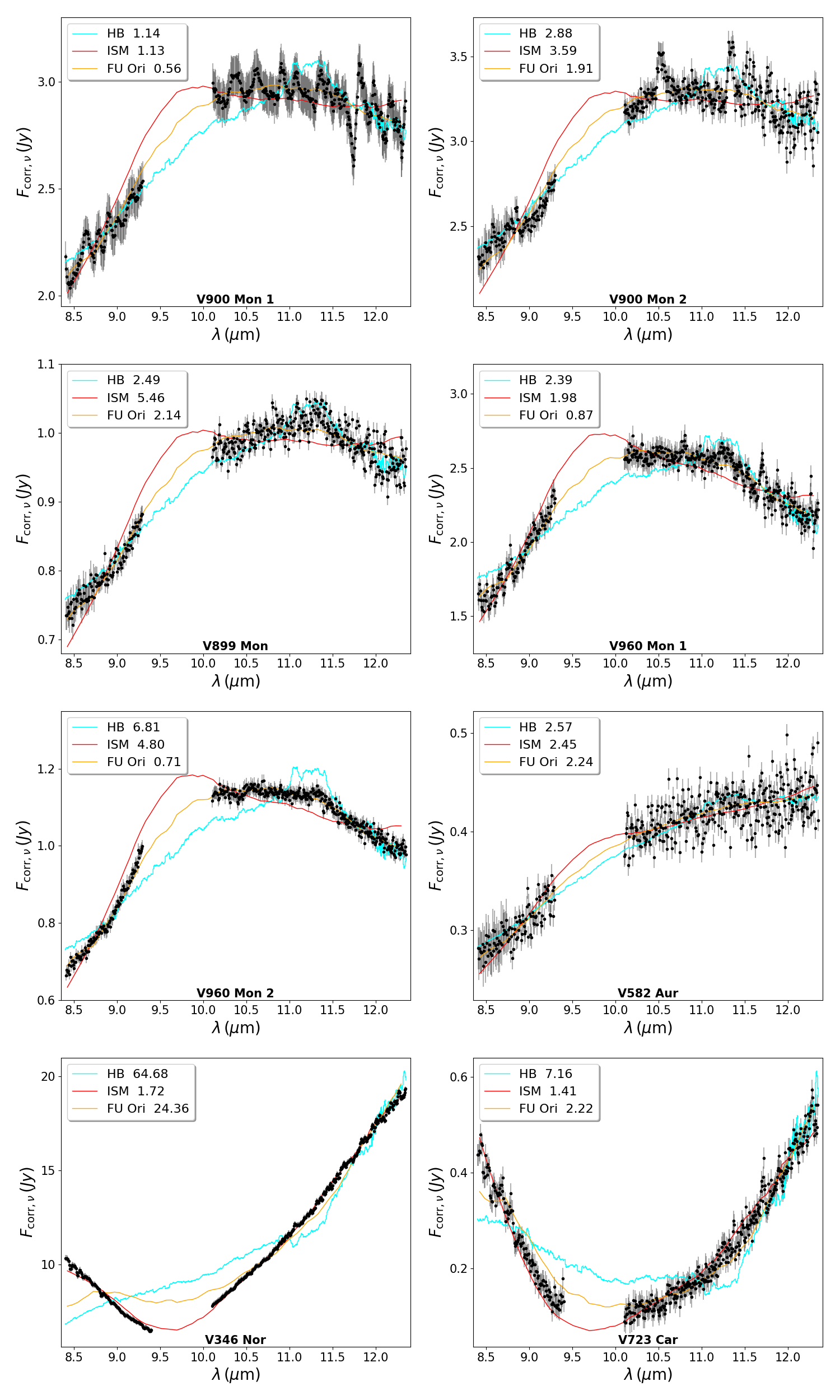}
   \caption{VLT/VISIR spectra of our targets (black dots with error bars) and V346 Nor \citep{kospal2020}. The colored curves are different templates fitted to the observed data points: the spectrum of comet Hale-Bopp (cyan, \citealt{hale-bopp_sp}), the interstellar dust grains (red, \citealt{kemper2004}) and FU Orionis (orange, \citealt{green2006}). The numbers in the legend show the $\chi^2$ values of the fits.\label{fig:jacob}}
\end{figure}

Fig.~\ref{fig:jacob} shows the VISIR spectra of our targets. We detected the 8$-$13$\,\mu$m emission of all five targets. Typical signal-to-noise ratios are 10$-$20 for V723~Car and 25$-$100 for the other targets. Out of our sample, one object (V723~Car) shows the 10$\,\mu$m silicate feature in absorption, while the remaining four (V582~Aur, V899~Mon, V900~Mon, and V960~Mon) have silicate in emission. Note that there is a significant difference in the absolute flux levels of our two V960~Mon spectra. This is due to the fact that the two spectra were taken 2.6 years apart and the source had been significantly fading in this period \citep{hackstein2015}.

The strength and shape of the silicate feature contains information about the composition and typical size of the grains that emit (or absorb) the feature. To characterize the observed spectral shapes, we fitted three different templates to the observed spectra: those of the interstellar matter (ISM), comet Hale-Bopp, and FU~Ori. The ISM typically contains submicron-sized amorphous silicates and displays a characteristic triangular shape that peaks at 9.7$\,\mu$m \citep{kemper2004}. Comets in the solar system may contain high fractions of crystalline olivines and pyroxenes (up to 90\%, e.g., \citealt{hale-bopp_sp,harker2002}), showing narrower peaks compared to the broad amorphous feature. FU~Ori, the prototype of the FUor class, exhibits a relatively weak emission feature with a plateau between 9.5$-$12$\,\mu$m, which is due to amorphous silicates that underwent significant grain growth \citep{przygodda2003, quanz2007c}.

We adopted a model fitting procedure similar to that used by \citet{varga2017} for the T Tauri-type star DG~Tau. We model the spectra as a combination of a linear continuum and one of the three template features which can be either in absorption or in emission. Emission features are optically thin, therefore the feature strength is scaled by a simple multiplicative factor during fitting. Absorption features can be optically thick, therefore in these cases the templates were scaled as $e^{-\tau(\lambda)}$. Fitting was restricted to the 8.4$-$9.4\,$\mu$m and 10.0$-$12.5\,$\mu$m wavelength ranges to avoid the ozone absorption and the edges of the atmospheric window. The best-fitting model of each type can be seen in Fig.~\ref{fig:jacob}. 

\subsection{Objects with silicate absorption}

The absorption feature of V723~Car seems to be most similar to the ISM template, suggesting that the absorption is caused by small amorphous dust grains. We calculated the optical depth at 9.7$\,\mu$m, and obtained $\tau_{9.7\mu\rm{m}}$ = 2.22. We converted it to $A_{9.7\mu\rm{m}}$, then to $A_V$ (adopting $A_V/A_{9.7\mu\rm{m}} = 18.5$ from \citealt{mathis1990}), and obtained $A_V = 45$\,mag. Using SED fitting, \citet{tapia2015} determined $A_V$ = 15$-$18\,mag circumstellar and 53$-$54\,mag interstellar extinction. Our value derived from the depth of the silicate absorption is in-between these numbers.

For comparison, we did a similar template fitting to the VISIR spectrum of V346~Nor, another embedded FUor \citep{kospal2020}. For this object, the best-fitting template is also that of the ISM. Here, we obtained $\tau_{9.7\mu\rm{m}}$ = 0.76 and $A_V$ = 15\,mag. The difference in absorption depth and shape between V723~Car and V346~Nor is most probably due to the different amounts of line-of-sight extinction leading to different optical depths in the silicate feature. The absorption in V723~Car is more optically thick than in V346~Nor, leading to almost zero fluxes at the bottom of the feature and a distortion of the typical triangular spectral shape.

\subsection{Objects with silicate emission}

The $\chi^2$ values in Fig.~\ref{fig:jacob} indicate that for all of the emission spectra, the FU~Ori template fits better than either the ISM or the Hale-Bopp template.
Therefore, we can conclude that in general the silicate emission features of our targets seem more similar to that of FU~Ori than that of the ISM or Hale-Bopp. We note that for V900~Mon, the first spectrum was taken at very high airmass, and is probably unreliable, as suggested also by the large wave-like artifacts especially between 10$-$12.5$\,\mu$m. Therefore, we do not interpret this spectrum further.

According to \citet{quanz2007c}, the Spitzer/IRS spectrum of FU~Ori can be fitted with a dust composition containing 57\% amorphous olivine and 43\% amorphous pyroxene. The model did not contain any crystalline forsterite, crystalline enstatite,  silica, or amorphous carbon. Most of the grains (84\%) in their model are $>1\,\mu$m in size. The similarity of our targets' spectra to that of FU~Ori suggests that these objects also contain almost exclusively amorphous silicates and grains must be mostly larger than micron-sized, but with varying amounts of submicron-sized grains present as well. We do not attempt a detailed mineralogical decomposition of our spectra; this would require higher signal-to-noise and calibration accuracy than our data possess.

\citet{vanboekel2003} and \citet{przygodda2003} found a correlation between the strength of the silicate feature (as defined by the maximum of the silicate feature over the continuum) and the ratio of the continuum-subtracted fluxes at 11.3$\,\mu$m over that at 9.8$\,\mu$m. They based this on a sample of a dozen Herbig Ae/Be stars \citep{vanboekel2003} and a dozen T Tauri stars \citep{przygodda2003} observed with the Thermal Infrared Multi Mode Instrument 2 (TIMMI2) at the 3.6\,m telescope at ESO's La Silla observatory in Chile and the Short Wavelength Spectrometer (SWS) onboard ESA's Infrared Space Observatory (ISO). They interpreted the correlation as the removal of small (0.1$\,\mu$m) grains from the disk surface while the large (1$-$2$\,\mu$m) grains remain.

\begin{figure}
   \includegraphics[width=\columnwidth,angle=0]{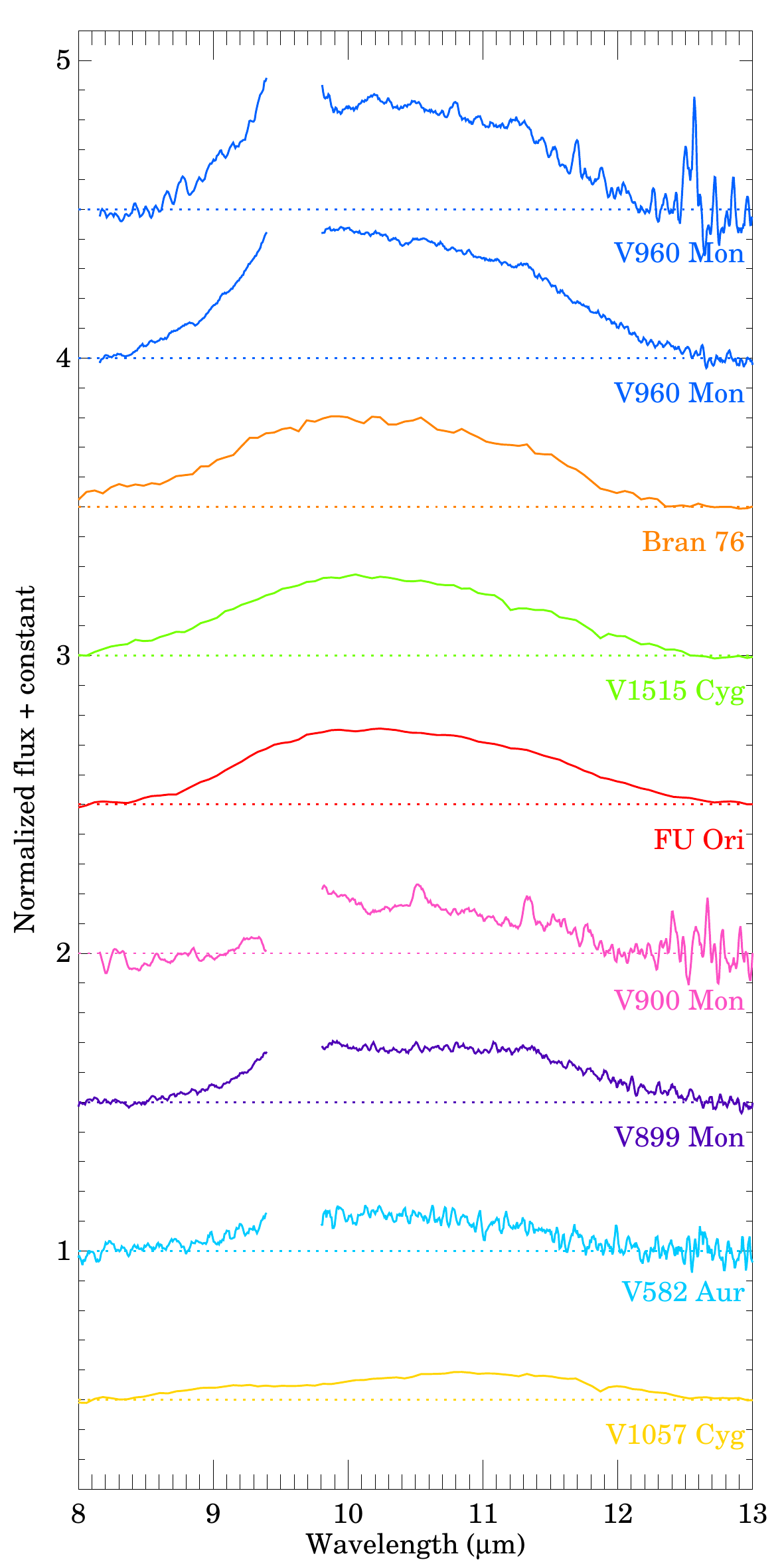}
   \caption{VLT/VISIR and Spitzer/IRS spectra of FUors with silicate emission feature. A linear continuum was  subtracted from the spectra, and they were also normalized by the average value of the fitted continuum. The spectra are plotted in order of decreasing feature strength from top to bottom. We smoothed the VISIR spectra using a 5-point wide moving window.\label{fig:normalized}}
\end{figure}

\begin{figure}
   \includegraphics[width=\columnwidth,angle=0]{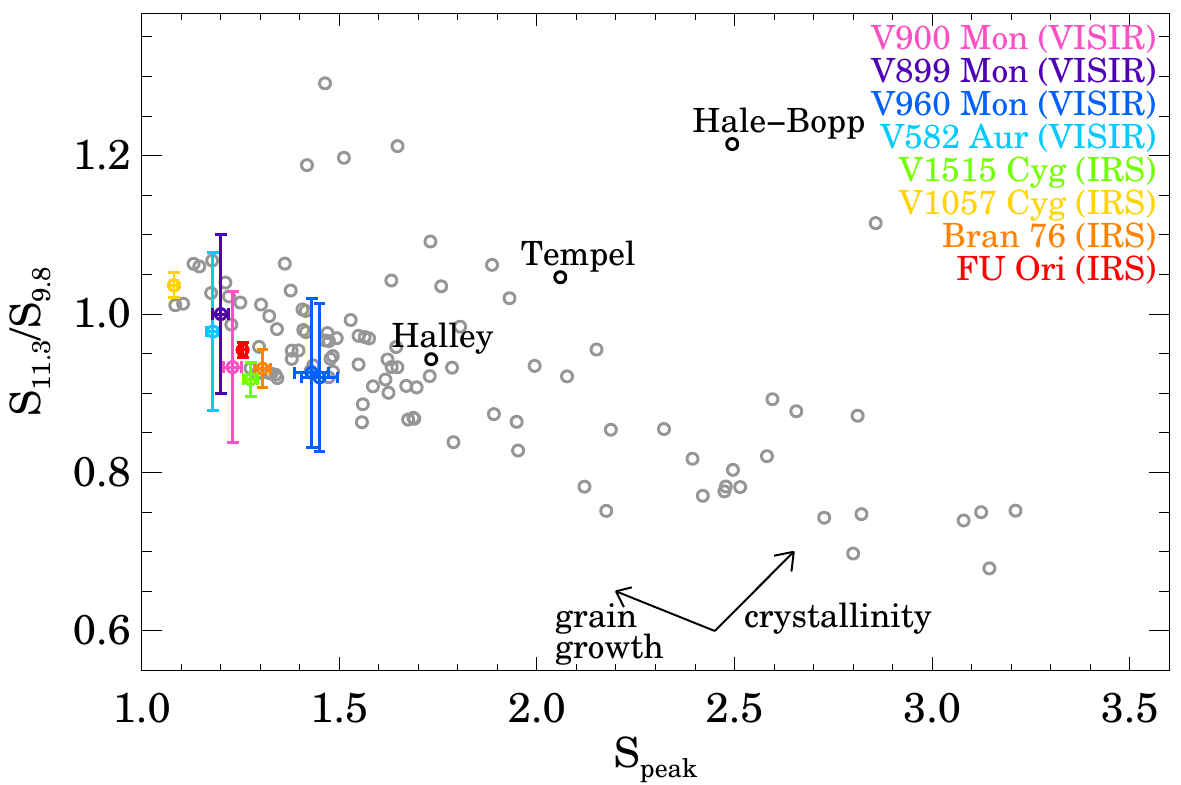}
   \caption{Ratio of the continuum subtracted and normalized 11.3/9.8$\,\mu$m flux ratio versus the feature strength. Gray dots mark Spitzer spectra of T Tauri and Herbig Ae/Be stars from the samples presented in \citet{przygodda2003}, \citet{kesslersilacci2006}, and \citet{olofsson2009}. Black dots show Solar System comets.\label{fig:przygodda}}
\end{figure}

These studies were later followed up using Spitzer/IRS spectra by \citet{kesslersilacci2006}, who confirmed that there are many weak and flat silicate features consistent with micron-sized grains indicating fast grain growth, and by \citet{olofsson2009}, who also found  micron-sized crystalline grains in many cases, in addition to the micron-sized amorphous grains. To check where our targets fall in the feature strength vs.~11.3/9.8$\,\mu$m flux ratio correlation, we first tried to reproduce the correlation for a merged sample of the targets from \citet{przygodda2003}, \citet{kesslersilacci2006}, and \citet{olofsson2009}, and downloaded their Spitzer/IRS spectra from the CASSIS database\footnote{https://cassis.sirtf.com/}. We fitted linear continua between 7.5$-$7.9$\,\mu$m and 12.9$-$13.1$\,\mu$m and followed the definitions of \citet{kesslersilacci2006} to calculate the flux ratios and feature strengths. We applied the same procedure to our VISIR spectra, except that we fitted the continua between 8.0$-$8.5$\,\mu$m and 12.5$-$13$\,\mu$m due to the restricted wavelength coverage of VISIR compared to IRS. The normalized spectra, in order of decreasing feature strength, are plotted in Fig.~\ref{fig:normalized}, while the resulting flux ratio vs.~feature strength graph can be seen in Fig.~\ref{fig:przygodda}. In this graph, we also indicated with arrows what direction a data point moves due to grain growth or a larger fraction of crystalline silicates, according to the modeling presented in \citet{kesslersilacci2006} and \citet{olofsson2009}.

The emission features we observed in our targets with VISIR are typically weak, broad, and flat. The strongest silicate emission feature ever observed in a FUor is displayed by V960~Mon. The two spectra of this star were taken with 2.6 years apart, during which time the source faded almost by a factor of two both in the continuum and in the feature. Remarkably, the relative strength and shape of the feature remained practically the same.

The weak and flat emission features in the FUors mean that these objects are clustered in the left side of the graph in Fig.~\ref{fig:przygodda}, indicating significant grain growth. Some of our targets are among the objects showing the weakest and flattest detectable silicate emission feature. The FUors are also located mostly at the lower edge of the area occupied by regular T Tauri and Herbig Ae/Be stars (gray dots), indicating that the grains around them are mostly amorphous. Nevertheless, there are some differences in feature strength between the individual FUors. The observed trend follows the same as in T Tauri and Herbig Ae/Be stars, i.e., a smaller $S_{\rm peak}$ is accompanied by an $S_{11.3}/S_{9.8}$ flux ratio closer to 1.0. As indicated in the figure, this is due to grain growth, suggesting that small grains contribute at varying degrees to the feature in the different FUors.

\section{Discussion}
\label{sec:dis}

In the following, we check whether the appearance of the silicate feature in absorption or emission  in our targets correlates with any of the system parameters, like the protostellar classification or the mass of the circumstellar material. We use our sample from Tab.~\ref{tab:log} together with V346~Nor in the following discussion.

There are three Class~I objects in the sample. V723~Car and V346~Nor show silicate absorption, while V900~Mon has weak silicate emission. The Class~II objects in the sample, V582~Aur, V899~Mon, and V960~Mon all show silicate emission. This suggests that, at least for embedded sources, the circumstellar geometry has an important role in determining the silicate feature. For instance, in case of a wide enough outflow cavity and favorable inclination, we can have a direct view of the disk, where silicate emission arises. Indeed, based on ALMA observations of the CO(2$-$1) rotational line, V900~Mon has an outflow opening angle of 70$^\circ$ on the blueshifted side and disk inclination of less than 48$^\circ$ from face-on \citep{takami2019}, while V346~Nor has an opening angle of only 40$^\circ$ on the blueshifted side of the outflow \citep{kospal2017c} and a disk inclination between 20$-$40$^\circ$ from face-on, as suggested by ALMA continuum emission (K\'osp\'al et al.~in prep.). This suggests that we observe V346~Nor through thick envelope material entrained in the outflow cavity walls, while we see V900~Mon through the cleared-out outflow cavity. We have no information on the circumstellar geometry of V723~Car other than that it has a very massive envelope (on the order of thousand solar masses, \citealt{tapia2015}), in line with the deep silicate absorption.

In conclusion, for our small sample of three Class~I objects, a silicate feature in absorption means that our line of sight is through the envelope, while a silicate feature in emission means that we see the object through the outflow cavity. The importance of circumstellar geometry and viewing angle on the appearance of the 10$\,\mu$m silicate feature was noticed earlier, both in model simulations \citep{robitaille2006} and in observations \citep{furlan2008}.

Based on interferometric observations of the millimeter continuum emission, the mass of the circumstellar matter can be estimated. Considering the spatial resolution provided by the synthesised beam (80--170\,au), this is most likely the mass of the disk for our targets. V582~Aur was observed at 3\,mm with the Plateau de Bure Interferometer and a total circumstellar mass of 0.04$\,M_{\odot}$ was calculated assuming optically thin emission at this wavelength \citep{feher2017}. ALMA continuum observations at 1.3\,mm are available for V899~Mon, V900~Mon,  V960~Mon, and V346~Nor (K\'osp\'al et al.~in prep.). By fitting radiative transfer models similar to those of \citet{cieza2018} to the observed brightness profiles, we found total masses of 0.03, 0.3, 0.2, and 0.2$\,M_{\odot}$, respectively. This suggests that, at least for eruptive young stellar objects, the circumstellar mass does not correlate one-to-one with either the protostellar class or the appearance of the silicate feature. The measured mass varies at least an order of magnitude within our Class~II objects, both Class~I and Class~II objects may have similar circumstellar masses, and both objects with silicate emission and absorption may have similar circumstellar masses.

Fig.~\ref{fig:przygodda} shows that normal, non-eruptive T Tauri and Herbig Ae/Be stars have a wide variety of silicate emission feature strengths and shapes, suggesting that some have small amorphous grains while others show varying degrees of grain growth and crystallization.  FUors show less variety: they all have amorphous, mostly large grains. This holds true not only for the objects whose VISIR spectra are presented here, but also for objects measured earlier with Spitzer/IRS, like FU~Ori, Bran~76, V1057~Cyg, and V1515~Cyg \citep{green2006}.

It is surprising to find large grains in disk surfaces as they should be settled in the midplane. \citet{vanboekel2003} proposed two possible explanations for this: turbulent mixing from the midplane, or a supply of large grains from the inner disk by an X-wind. These mechanisms may operate in FUors at an elevated level due to the on-going outburst. In a large sample of classical T Tauri stars, \citet{watson2009} found that increasing crystallinity is linked to dust settling to the disk midplane, but found no correlation between the crystalline mass fraction and stellar mass or luminosity, stellar accretion rate, disk mass, or disk/star mass ratio. In this context, although FUors have high accretion rates and large luminosities, they may lack crystalline silicates due to the dust not being settled in the midplane in their disks.

To learn more about the distribution, composition, and other properties of silicate grains in FUors, two new possibilities are open. MATISSE, the second generation infrared interferometer on the VLT is able to resolve the emitting region in the $N$ band down to spatial scales of a few milli-arcsecond for the brighter sources. MATISSE has the potential to solve the puzzling lack of crystalline silicates in FUor disks by separating the emission from the innermost, hottest disk regions. The MIRI instrument of the JWST will be able to detect sources down to the $\mu$Jy level at 10$\,\mu$m. The JWST's superior sensitivity and stability makes it the perfect instrument for monitoring young eruptive stars. It is expected to detect subtle changes in the silicate feature during the episodic accretion events, so that we can learn how these events affect the grains in the planet-forming region.


\acknowledgments

This project has received funding from the European Research Council (ERC) under the European Union's Horizon 2020 research and innovation programme under grant agreement No 716155 (SACCRED).

\vspace{5mm}
\facilities{VLT:Melipal, Spitzer}



\end{document}